# Fidelity, purity and entanglement of two-mode spatially Gaussian-entangled light fields in Turbulence Atmosphere

Li-Gang Wang, [1, 2] and Shi-Yao Zhu, [1, 2, 3]

[1] *Center of Optics, Department of Physics, The Chinese University of Hong Kong, Shatin, Hong Kong*

[2] *Department of Physics, Zhejiang University, Hangzhou 310027, China*

[3] *Department of Physics, The Hong Kong Baptist Univesity, Kowloon Tong, Hong Kong*

In this paper, we investigate the propagation of two-mode spatially Gaussian-entangled quantum light fields passing through the turbulence atmosphere. From the propagation formula of the two-mode wave function in the position representation, we have derived the analytical expressions for the fidelity, purity and logarithmic negativity (entanglement) of the resulting quantum state after the long-distance atmospheric transportation. Based on the derived formulae, the effects of the atmospheric turbulences on the evolutions of quantum properties of the resulting two-mode quantum state are discussed in detail under different input parameters of the initial two-mode quantum state. The results show that the maximal distributing distance $L$ of quantum entanglement is strongly dependent on the atmospheric conditions: when the atmospheric turbulence becomes stronger and stronger, the maximal distance $L$ becomes shorter and shorter, and both the fidelity and purity decrease quicker and quicker as functions of propagating distances. Under a certain atmospheric condition, with the increasing of the input entanglement of the initial two-mode spatially Gaussian-entangled quantum state, the maximal distributing distance for preserving the entanglement gradually increases and always has a saturated (upper) limitation, and both the evolutions of the fidelity and purity are affected by the input parameters of the initial two-mode quantum state, Finally the optimal parameters of the input two-mode quantum state with the fixed input entanglement are discussed in order to obtain the optimal transfer distribution of the quantum entanglement over a long distance under a certain atmosphere. Our theoretical results are very helpful for building the distribution of the quantum entanglement via free-space atmosphere link.

PACS numbers: 03.67. Hk, 42.68.Bz, 03.67.Lx

## I. Introduction

The long-distance distribution of quantum entanglement between two distant observers has become a hot topic in quantum communication. [1-5] There are many experimental proposals on the long-distance quantum communication with entangled photons. The best-known examples are quantum teleportation [6] and quantum key distribution. [7] Quantum repeater [8] and quantum relay [9] are invented for increasing the maximal distance an entangled quantum state can be distributed over. Recently optical fiber [10] and free-space links have been the excellent candidates for low-loss distribution of the entangled quantum state over long distances. Aspelmeyer *et al.* [2] have proposed to use satellite-based free space distribution for single photons or entangled photon pairs over global distances. Resch *et al.* [11] have reported that the polarization-entangled photon pairs from parametric down-conversion have been successfully distributed over 7.8 km through the intra-city atmosphere at night. Later Pan's group [3] further reported a significant step towards satellite-based global quantum communication by showing the free-space distribution of entangled photon pairs over a noisy ground atmosphere of 13 km. Meanwhile, Marcikic *et al.* [12] have experimentally implemented the distribution of time-bin entangled qubits over 50 km of optical fibers, and Takesue *et al.* [13] carried out the differential phase shift quantum key distribution experiment over 105 km fiber. Most recently Ursin *et al.* [5] successfully demonstrated entanglement-based quantum key distribution over 144 km between two Canary Islands La Palma and Tenerife, and their experiment is an essential step towards future satellite-based distribution of quantum entanglement, to establish a worldwide network for quantum communication. [14]

In the theoretical aspect, Waks *et al.* [15] have analyzed the effects of the absorptive losses and the dark count of the detectors on the limitation of the distance for distributing entanglement about the order of 100 km. Up to now, there are still no theoretical investigation on how the atmosphere effect limits on the distribution of quantum entanglement in detail. To explore entanglement, there are a number of degrees of freedom to choose from momentum, time-bins, [12] polarization, orbital angular momentum, [16] and transverse position-momentum. [17-18] In this paper, we consider the propagation of the two-mode spatially Gaussian-entangled light fields passing through the turbulent atmosphere. In our consideration, we focus on the propagation properties of two-mode spatially Gaussian-entangled light fields in a turbulence atmosphere, and analyze how the atmospheric turbulence affects on the quantum properties of the output quantum state after a long-distance transportation. It should be pointed out that, although our investigation is on the distribution of the transverse spatial entanglement, the results might be profitable for the other types of entanglement transportation in atmosphere. The paper is organized as follows: in Sec. II, we describe our model on the two-mode spatially Gaussian-entangled quantum light fields and their propagations, and we derive the formulae of the fidelity, purity and logarithmic negativity of the output two-mode quantum state after the distant atmospheric transportation; in Sec. III, based on the derived analytical expressions, we numerically demonstrate how the atmospheric turbulence affects on the quantum properties of the output two-mode quantum state in detail; and the final conclusion could be found in Sec. IV.

**II. Theory**

As well known, in wave mechanics, the fundamental physical entities are "particles" always described by a wave function, which is equivalent to the fields in quantum field theory. According to the one-to-one correspondence between *modes* in quantum field theory and *states* in wave mechanics, Smith and Raymer [19] have introduced a two-photon wave function and its equation of motion in the position representation. Saleh *et al.* [20] have verified that the two-photon probability amplitude, describing light fields in a two-photon entangled state, obeys equations identical to the Wolf equations. These theoretical developments give us the opportunities to investigate the propagation properties of two-mode (or two-photon) quantum state in the position representation. For the two-mode (or two-photon) quantum state $|\Psi\rangle$ generated from the spontaneous parametric down-conversion or the two-mode correlated lasers in the momentum space is given by

$$|\Psi\rangle = \int dp_1 \int dp_2 \tilde{\psi}(p_1, p_2) | p_1 \rangle | p_2 \rangle \tag{1a}$$

and its equivalent form in the position space is given by

$$|\Psi\rangle = \int dx_1 \int dx_2 \psi(x_1, x_2) | x_1 \rangle | x_2 \rangle, \tag{1b}$$

where $\tilde{\psi}(p_1, p_2)$ and $\psi(x_1, x_2)$ are the two-mode wave functions (or probability amplitudes) in the momentum and position representations, respectively, and $x_j$ and $p_j$, respectively, are the transverse position and the transverse momentum of the photon in mode $j = 1, 2$. This two-mode state is entangled if $\psi(x_1, x_2)$ is inseparable.

Here we consider a two-mode spatially Gaussian-entangled light field passing through a turbulence atmosphere, as shown in Fig. 1, and the two modes may be with different wavelengths $\lambda_1$ and $\lambda_2$. The two-mode wave function for the initial two-mode spatially Gaussian-entangled light field $|\Psi_{in}\rangle$ at the input planes ($z_1 = 0$ and $z_2 = 0$) are assumed to be the Gaussian form as follows,

$$\psi_{in}(x_1, x_2) = G_0 \exp[-\frac{x_1^2}{4\sigma_p^2} - \frac{x_2^2}{4\sigma_p^2} - \frac{(x_1 - x_2)^2}{2\sigma_c^2}], \tag{2}$$

where the normalized factor $G_0^2 = \frac{2}{\pi\Delta^2}$ with $\frac{1}{\Delta^4} = \frac{1}{\xi^4} - \frac{1}{4\sigma_c^4}$ and $\frac{1}{\xi^2} = \frac{1}{4\sigma_p^2} + \frac{1}{2\sigma_c^2}$, and $\sigma_p^2$ denotes the quantum fluctuations of $<\Delta x_1^2>$ and $<\Delta x_2^2>$ for a product quantum state (i. e., when $\sigma_c \to \infty$), and the parameter $\sigma_c$ denotes the transverse spatial quantum correlation between two modes. In the limit of $\sigma_p \to \infty$ and $\sigma_c \to 0$, Eq. (2) becomes $\psi_{in}(x_1, x_2) = G_0 \delta(x_1 - x_2)$, which corresponds to the original Einstein-Podolsky-Rosen (EPR) state. [21] For the fixed value $\sigma_p$, the smaller $\sigma_c$, the higher quantum correlation; while for the fixed value $\sigma_c$, the larger $\sigma_p$, the higher quantum correlation. Therefore for the practical situations with any finite values $\sigma_p$ and $\sigma_c$, both $\sigma_p$ and $\sigma_c$ are related with the quantifications of the quantum entanglement. As we shown in the below, the logarithmic negativity of the entanglement for the input quantum state of Eq. (2) actually could be expressed as $E_N = \frac{1}{2}\ln(1 + 4W^2)$, where $W = \sigma_p / \sigma_c$ determines the initial quantum entanglement. We emphasize here that the quantum state by Eq. (1b) with the two-mode wave function by Eq. (2) describes *a kind of two-mode transverse spatially Gaussian-entangled quantum light fields.*

The propagation formula of the wave function in the position representation for any two-mode quantum light fields between the input and output planes is given by [20]

$$\psi_{out}(u_1, u_2) = \iint \psi_{in}(x_1, x_2) G_1(u_1, x_1) G_2(u_2, x_2) dx_1 dx_2, \tag{3}$$

where $G_1(u_1, x_1)$ and $G_2(u_2, x_2)$ are the Green's functions from the input planes $x_1$ ($x_2$) to the output planes $u_1$ ($u_2$) in paths 1 (2), respectively. Under the paraxial approximation, the Green's functions $G_1(u_1, x_1)$ and $G_2(u_2, x_2)$ for a turbulent atmosphere could be approximately written as [22]

$$G_j(u_j, x_j) = (-\frac{i}{\lambda_j z_j})^{1/2} e^{ik_j z_j} e^{\frac{ik_j}{2z_j}(x_j-u_j)^2 + \varphi_j(x_j, u_j)}, \tag{4}$$

where $j = 1, 2$ denote the two paths (see Fig. 1), and the wave number $k_j = 2\pi/\lambda_j$ and the function $\varphi_j(x_j, u_j)$ represents the random complex phase of a light field due to the propagation in a turbulent atmosphere. Using Eq. (4), we can express Eq. (3) as follows:

$$\psi_{\text{out}}(u_1, u_2) = \frac{ie^{i(k_1 z_1 + k_2 z_2)}}{(\lambda_1 \lambda_2 z_1 z_2)^{1/2}} \iint \psi_{\text{in}}(x_1, x_2) e^{\frac{ik_1}{2z_1}(x_1-u_1)^2 + \varphi_1(x_1, u_1)} e^{\frac{ik_2}{2z_2}(x_2-u_2)^2 + \varphi_2(x_2, u_2)} dx_1 dx_2. \tag{5}$$

Equation (5) describes a general propagating integral equation of the output two-mode wave function for the two-mode spatially entangled quantum light fields passing through the turbulence atmosphere. In fact, as the entangled quantum light fields propagate through the atmosphere, due to the effect of the random factor $\varphi_j(x_j, u_j)$, the output quantum state $|\Psi_{\text{out}}\rangle$ with the output wave function of Eq. (5) actually becomes a mixed two-mode quantum state. The propagating properties of the output two-mode quantum state $|\Psi_{\text{out}}\rangle$ could be found by taking the ensemble average over the observable quantities, thus the probability density distribution of the output two-mode spatially entangled light fields could be expressed as $P(u_1, u_2) = \langle |\psi_{\text{out}}(u_1, u_2)|^2 \rangle_e$, where the symbol $\langle \rangle_e$ denotes the ensemble average. From Eq. (5), the probability density distribution could be written as

$$P(u_1, u_2) = \frac{1}{\lambda_1 \lambda_2 z_1 z_2} \iiiint dx_1 dx_2 dx_3 dx_4 \psi_{in}^*(x_1, x_2) \psi_{in}(x_3, x_4) e^{-\frac{ik_1}{2z_1}(x_1-u_1)^2} e^{-\frac{ik_2}{2z_2}(x_2-u_2)^2}$$
$$\times e^{\frac{ik_1}{2z_1}(x_3-u_1)^2} e^{\frac{ik_2}{2z_2}(x_4-u_2)^2} \langle e^{\varphi_1^*(x_1,u_1)+\varphi_1(x_3,u_1)} \rangle_e \langle e^{\varphi_2^*(x_2,u_2)+\varphi_2(x_4,u_2)} \rangle_e. \tag{6}$$

In the above derivations, we have used the assumption that the statistical properties of the atmospheric turbulences in the two paths are independent of each other. The statistical ensemble average on the random complex phase for the homogeneous atmospheric turbulence can be approximated by [23]

$$\langle e^{\varphi_j^*(x_1,u_1)} e^{\varphi_j(x_2,u_2)} \rangle_e = e^{-\frac{1}{2} D_{\varphi_j}(x_1,u_1;x_2,u_2)}$$
$$\cong e^{-\frac{1}{\rho_j^2}[(x_1-x_2)^2 + (x_1-x_2)(u_1-u_2) + (u_1-u_2)^2]}, \tag{7}$$

where $j = 1, 2$. Here $D_{\varphi_j}$ is the structure function of the random complex phase in Rytov's representation, and $\rho_j = (0.545 C_n^2 k_j^2 z)^{-3/5}$ is the coherence length of a spherical wave propagating in atmospheric turbulence characterized by the refractive index structure parameter $C_n^2$. It should be mentioned that Eq. (7) is valid in both the weak and strong atmospheric fluctuations. [24] Substituting Eqs. (2) and (7) into Eq. (6), after the tedious calculations, the resulting probability density distribution is then given by

$$P(u_1, u_2) = \frac{G_0^2}{\sqrt{Q(z_1, z_2)}} e^{-(\alpha u_1^2 + \beta u_1^2 + 2\gamma u_1 u_2)}, \tag{8}$$

where $\alpha = [\frac{2}{\xi^2} + \frac{2\lambda_2^2 z_2^2}{\pi^2 \Delta^4}(\frac{2}{\rho_2^2} + \frac{1}{\xi^2})]/Q(z_1, z_2)$, $\beta = [\frac{2}{\xi^2} + \frac{2\lambda_1^2 z_1^2}{\pi^2 \Delta^4}(\frac{2}{\rho_1^2} + \frac{1}{\xi^2})]/Q(z_1, z_2)$,

$\gamma = (\frac{\lambda_1 z_1 \lambda_2 z_2}{\pi^2 \Delta^4} - 1)/[Q(z_1, z_2)\sigma_c^2]$, and $Q(z_1, z_2) = 1 + \frac{\lambda_1 z_1 \lambda_2 z_2}{2\pi^2 \sigma_c^4} + \sum_{j=1}^{2} \frac{\lambda_j^2 z_j^2}{\pi^2 \xi^2}\left(\frac{2}{\rho_j^2} + \frac{1}{\xi^2}\right)$

$+ \frac{\lambda_1^2 z_1^2 \lambda_2^2 z_2^2}{\pi^4 \Delta^8}[1 + \frac{4\Delta^4}{\rho_1^2 \rho_2^2} + \frac{2\Delta^4}{\xi^2}(\frac{1}{\rho_1^2} + \frac{1}{\rho_2^2})]$. From Eq. (8), it is clear found that the probability density distribution of the output two-mode quantum state is strongly affected by the atmospheric turbulence with the increasing of the propagating distance and it diffuses much wider than that in the free space, and the coefficient $\gamma$ is gradually decreasing much faster than other two coefficients $\alpha$ and $\beta$ due to the effect of the atmospheric turbulence. Therefore the correlation (including quantum and classical correlations) of the output quantum state is qualitatively decreasing due to the atmospheric turbulence.

In order to quantify the quantum properties of the two-mode spatially entangled state passing through the atmosphere, let us first calculate the fidelity of the output two-mode quantum state $|\Psi_{out}\rangle$ relative to the input quantum state $|\Psi_{in}\rangle$, which is given by [25]

$$F = \left\langle \left| \langle \Psi_{out} | \Psi_{in} \rangle_Q \right| \right\rangle_e = \left\langle \left| \iint du_1 du_2 \psi_{out}^*(u_1, u_2) \psi_{in}(u_1, u_2) \right| \right\rangle_e, \tag{9}$$

where $\langle \rangle_Q$ denotes the quantum mechanical inner product, and $\langle \rangle_e$ denotes the ensemble average. The fidelity $F$ indicates the distinguishability of the two states and $F^2$ denotes the probability of the system being in $|\Psi_{out}\rangle$ after the measurement. [25] Substituting Eqs. (2) and (5) into Eq. (9) and using Eq. (7), we obtain the analytical expression as follows:

$$F = [1 + \frac{3\Delta^4}{\xi^2}(\frac{1}{\rho_1^2} + \frac{1}{\rho_2^2}) + \frac{9\Delta^4}{\rho_1^2 \rho_2^2}]^{-1/4} \left\{ 1 + \frac{\lambda_1 z_1 \lambda_2 z_2}{8\pi^2 \sigma_c^4} + \frac{1}{4\pi^2 \xi^2}[\lambda_1^2 z_1^2(\frac{1}{\rho_1^2} + \frac{1}{\xi^2}) \right.$$
$$\left. + \lambda_2^2 z_2^2(\frac{1}{\rho_2^2} + \frac{1}{\xi^2})] + \frac{\lambda_1^2 z_1^2 \lambda_2^2 z_2^2}{16\pi^2 \Delta^8}[1 + \frac{3\Delta^4}{\xi^2}(\frac{1}{\rho_1^2} + \frac{1}{\rho_2^2}) + \frac{9\Delta^4}{\rho_1^2 \rho_2^2}] \right\}^{-1/4}. \tag{10}$$

Equation (10) shows that the atmospheric turbulence has strongly affected on the fidelity of the output two-mode quantum state. In the vacuum, Eq. (10) could be simplified into $F = (1 + \frac{\lambda_1 z_1 \lambda_2 z_2}{8\pi^2 \sigma_c^4} + \frac{\lambda_1^2 z_1^2 + \lambda_2^2 z_2^2}{4\pi^2 \xi^4} + \frac{\lambda_1^2 z_1^2 \lambda_2^2 z_2^2}{16\pi^2 \Delta^8})^{-1/4}$ by taking $\frac{1}{\rho_1^2} \to 0$ and $\frac{1}{\rho_2^2} \to 0$ (i. e., $C_n^2 \to 0$). Comparing Eq. (10) with the case in the vacuum, it is seen that, when the propagating distances increases, the fidelity of the output two-mode quantum state in the atmosphere decreases much faster than that through the free space (see the detailed discussion in the next section).

Now let us turn to evaluate the purity and entanglement of the two-mode spatially entangled quantum state passing through the atmosphere space, we use the method of the second moments of the canonical operators, proposed by Rendell and Rajagopal, [26] to calculate the $4\times 4$ covariance matrix $\mathbf{V}$. [26-29] For a two-mode Gaussian state, the $4\times 4$ covariance matrix $\mathbf{V}$ contains all the necessary information to determine its entanglement. The covariance matrix $\mathbf{V}$ could be written in terms of the three $2\times 2$ partitioned matrices $\mathbf{A}$, $\mathbf{B}$, $\mathbf{C}$ as follows [26]

$$\mathbf{V}=\begin{pmatrix} \mathbf{A} & \mathbf{C} \\ \mathbf{C}^T & \mathbf{B} \end{pmatrix}, \quad (11)$$

where $\mathbf{A}=\begin{pmatrix} \langle\langle\hat{u}_1^2\rangle_Q\rangle_e & \langle\langle\{\hat{u}_1\hat{p}_1\}\rangle_Q\rangle_e \\ \langle\langle\{\hat{u}_1\hat{p}_1\}\rangle_Q\rangle_e & \langle\langle\hat{p}_1^2\rangle_Q\rangle_e \end{pmatrix}$ and $\mathbf{B}=\begin{pmatrix} \langle\langle\hat{u}_2^2\rangle_Q\rangle_e & \langle\langle\{\hat{u}_2\hat{p}_2\}\rangle_Q\rangle_e \\ \langle\langle\{\hat{u}_2\hat{p}_2\}\rangle_Q\rangle_e & \langle\langle\hat{p}_2^2\rangle_Q\rangle_e \end{pmatrix}$, respectively, refers to the light modes in the paths 1 and 2, and $\mathbf{C}=\begin{pmatrix} \langle\langle\hat{u}_1\hat{u}_2\rangle_Q\rangle_e & \langle\langle\hat{u}_1\hat{p}_2\rangle_Q\rangle_e \\ \langle\langle\hat{p}_1\hat{u}_2\rangle_Q\rangle_e & \langle\langle\hat{p}_1\hat{p}_2\rangle_Q\rangle_e \end{pmatrix}$ describes the correlation between the two modes. Here $\hat{p}_j=-i\partial/\partial u_j$ is the transverse moment operator conjugate to the transverse position operator $\hat{u}_j$, $j=1,2$, and $\langle\{\hat{a}\hat{b}\}\rangle_Q=\langle(\hat{a}\hat{b}+\hat{b}\hat{a})\rangle_Q/2$. All the elements of $\mathbf{A}$, $\mathbf{B}$ and $\mathbf{C}$ are explicitly calculated by using $\langle\langle\hat{O}\rangle_Q\rangle_e \equiv \langle\iint du_1 du_2 \psi_{\text{out}}^*(u_1,u_2)\hat{O}\psi_{\text{out}}(u_1,u_2)\rangle_e$ and can be analytically expressed as

$$\langle\langle\hat{u}_1^2\rangle_Q\rangle_A = \frac{\Delta^4}{4\xi^2}+\frac{\lambda_1^2 z_1^2}{4\pi^2}\left(\frac{1}{\xi^2}+\frac{2}{\rho_1^2}\right), \quad \langle\langle\hat{u}_2^2\rangle_Q\rangle_A = \frac{\Delta^4}{4\xi^2}+\frac{\lambda_2^2 z_2^2}{4\pi^2}\left(\frac{1}{\xi^2}+\frac{2}{\rho_2^2}\right), \quad (12a)$$

$$\langle\langle\hat{p}_1^2\rangle_Q\rangle_A = \frac{1}{\xi^2}+\frac{6}{\rho_1^2}, \quad \langle\langle\hat{p}_2^2\rangle_Q\rangle_A = \frac{1}{\xi^2}+\frac{6}{\rho_2^2}, \quad (12b)$$

$$\langle\langle\{\hat{u}_1\hat{p}_1\}\rangle_Q\rangle_A = \frac{\lambda_1 z_1}{2\pi}\left(\frac{1}{\xi^2}+\frac{3}{\rho_1^2}\right), \quad \langle\langle\{\hat{u}_2\hat{p}_2\}\rangle_Q\rangle_A = \frac{\lambda_2 z_2}{2\pi}\left(\frac{1}{\xi^2}+\frac{3}{\rho_2^2}\right), \quad (12c)$$

$$\langle\langle\hat{u}_1\hat{u}_2\rangle_Q\rangle_A = \frac{\pi^2\Delta^4-\lambda_1 z_1 \lambda_2 z_2}{8\pi^2\sigma_c^2}, \quad \langle\langle\hat{p}_1\hat{p}_2\rangle_Q\rangle_A = -\frac{1}{2\sigma_c^2}, \quad (12d)$$

$$\langle\langle\hat{p}_1\hat{u}_2\rangle_Q\rangle_A = -\frac{\lambda_1 z_1}{4\pi\sigma_c^2}, \quad \langle\langle\hat{u}_1\hat{p}_2\rangle_Q\rangle_A = -\frac{\lambda_2 z_2}{4\pi\sigma_c^2}. \quad (12e)$$

In the derivations of Eqs. (12), we have used Eq. (7) and the following two equalities:

$$\left\langle e^{\varphi_j^*(x_1,u_1)}\frac{\partial}{\partial u_2}e^{\varphi_j(x_2,u_2)}\right\rangle_e = \frac{\partial}{\partial u_2}\left\langle e^{\varphi_j^*(x_1,u_1)}e^{\varphi_j(x_2,u_2)}\right\rangle_e, \quad (13a)$$

$$\left\langle e^{\varphi_j^*(x_1,u_1)} \frac{\partial^2}{\partial u_2^2} e^{\varphi_j(x_2,u_2)} \right\rangle_e = \frac{\partial^2}{\partial u_2^2} \left\langle e^{\varphi_j^*(x_1,u_1)} e^{\varphi_j(x_2,u_2)} \right\rangle_e, \qquad (13b)$$

where $j = 1,2$. It should be pointed out that all the first moments of the operators $\hat{u}_j$ and $\hat{p}_j$ are equal to zero, which do not affect any quantity related to entanglement or mixedness. From Eq. (11) and (12), one can easily prove that the inequality

$$\mathbf{V} + \frac{i}{2}\mathbf{\Omega} \geq 0 \qquad (14)$$

always holds, where $\mathbf{\Omega}$ is the standard symplectic form

$$\mathbf{\Omega} \equiv \begin{pmatrix} \Lambda & 0 \\ 0 & \Lambda \end{pmatrix}, \ \Lambda \equiv \begin{pmatrix} 0 & 1 \\ -1 & 0 \end{pmatrix}.$$

Actually, inequality (14) denotes the Heisenberg uncertainty principle. [27] Inequality (14) actually indicates that *the wave functions given by Eqs. (2) and (5) are the quantum fields not the classical fields.*

In the same time, from Eqs. (11) and (12), the purity of the output two-mode quantum state could be calculated by

$$\mu = \frac{1}{4\chi^{1/2}}, \qquad (15)$$

where

$$\chi = \text{Det}[\mathbf{V}] = \frac{1}{16}\left\{1 + \frac{6\Delta^4}{\xi^2}\left(\frac{1}{\rho_1^2} + \frac{1}{\rho_2^2}\right) + \frac{36\Delta^4}{\rho_1^2\rho_2^2} + \frac{4\lambda_1^2 z_1^2 \lambda_2^2 z_2^2}{\pi^4 \rho_1^2 \rho_2^2}\left[\frac{1}{\Delta^4} + \frac{3}{2\xi^2}\left(\frac{1}{\rho_1^2} + \frac{1}{\rho_2^2}\right) + \frac{9}{4\rho_1^2\rho_2^2}\right] \right.$$
$$\left. + \frac{2\Delta^4}{\pi^2\xi^2}\left[\frac{9\xi^2 \lambda_1 z_1 \lambda_2 z_2}{4\sigma_c^4 \rho_1^2 \rho_2^2} + \left(\frac{\lambda_1^2 z_1^2}{\rho_1^2} + \frac{\lambda_2^2 z_2^2}{\rho_2^2}\right)\left(\frac{1}{\Delta^4} + \frac{9}{\rho_1^2\rho_2^2}\right) + \frac{3}{2\xi^2}\left(\frac{\lambda_1^2 z_1^2}{\rho_1^4} + \frac{\lambda_2^2 z_2^2}{\rho_2^4} + \frac{4(\lambda_1^2 z_1^2 + \lambda_2^2 z_2^2)}{\rho_1^2 \rho_2^2}\right)\right]\right\}.$$

(16)

The purity is quantifying the degree of the mixedness of the output two-mode quantum state due to the effect of the turbulence atmosphere. From Eq. (16), it is found that, in the case of the vacuum ($\frac{1}{\rho_1^2} \to 0$ and $\frac{1}{\rho_2^2} \to 0$) the value $\chi$ is always a constant equal to $1/16$ and then the purity of the output quantum state is always equal to one, which indicates *the output quantum state is a pure state in the vacuum.* However, in the turbulent atmosphere, the output quantum state becomes a mixed quantum state, and then *the purity decreases as the increasing of the propagating distances.* Meanwhile, from Eq. (16), it is clear shown that the quantity $\chi$ increases with the increasing of the propagating distances due to the atmospheric effect, which indicates that the output quantum state is no more a pure state.

Now we consider the evolution of the entanglement for the two-mode spatially Gaussian-entangled light fields propagating through the atmosphere. As we know that inequality (14) can be recast as a constraint on the symplectic eigenvalues $n_{\mp}$ of the covariance matrix $\mathbf{V}$, [30] where $2n_{\mp}^2 = \eta(\mathbf{V}) \mp [\eta^2(\mathbf{V}) - 4\chi]^{1/2}$ and $\eta(\mathbf{V}) = \mathrm{Det}[\mathbf{A}] + \mathrm{Det}[\mathbf{B}] + 2\mathrm{Det}[\mathbf{C}]$. From Eq. (11) and (12), we can also easily prove the minimal symplectic eigenvalue satisfying the inequality $n_- \geq 1/2$. For two-mode Gaussian states, the necessary and sufficient separability criterion is positivity of the partially transposed matrix $\tilde{\mathbf{V}}$ of the covariance matrix $\mathbf{V}$ (the so-called "PPT criterion"). [27] According to Refs. [30-31], the symplectic eigenvalues $\tilde{n}_{\mp}$ of the partially transposed matrix $\tilde{\mathbf{V}}$ read

$$\tilde{n}_{\mp} = \left\{ \frac{\tilde{\eta}(\mathbf{V}) \mp [\tilde{\eta}^2(\mathbf{V}) - 4\chi]^{1/2}}{2} \right\}^{1/2}, \tag{17}$$

where

$$\tilde{\eta}(\mathbf{V}) = \mathrm{Det}[\mathbf{A}] + \mathrm{Det}[\mathbf{B}] - 2\mathrm{Det}[\mathbf{C}]$$
$$= \frac{1}{2}\left[1 + \frac{\Delta^4}{2\sigma_c^4} + \frac{3\Delta^4}{\xi^2}\left(\frac{1}{\rho_1^2} + \frac{1}{\rho_2^2}\right) + \frac{1}{2\pi^2}\sum_{j=1}^{2}\frac{\lambda_j^2 z_j^2}{\rho_j^2}\left(\frac{3}{\rho_j^2} + \frac{2}{\xi^2}\right)\right]. \tag{18}$$

As for the quantification of entanglement, there is still no full satisfactory method known at present for arbitrary mixed two-mode Gaussian states. Fortunately the logarithmic negativity turns out to be an entanglement monotone. From Eq. (17), it is easy to verify that the symplectic eigenvalue $\tilde{n}_+$ is always greater than 1/2 at any parameter condition, and has no effect on establishing the nonseparability of the quantum state. [30] The other symplectic eigenvalue $\tilde{n}_-$ may be less than 1/2 and determines the nonseparability of the output two-mode spatially Gaussian-entangled state. Therefore the logarithmic negativity can be a simple function of $\tilde{n}_-$ as follows [31]

$$E_N(\mathbf{V}) = \max\{0, -\ln[2\tilde{n}_-]\}. \tag{19}$$

This is a decreasing function of the symplectic eigenvalue $\tilde{n}_-$, and hence the symplectic eigenvalue $\tilde{n}_-$ completely quantifies the entanglement of the output two-mode spatially Gaussian-entangled quantum state. From Eqs. (16-19), it is easy to obtain the logarithmic negativity of the input two-mode quantum state given by $E_N = \frac{1}{2}\ln(1+4W^2)$, where the ratio $W = \sigma_p/\sigma_c$ determines the initial quantum entanglement. When $W = 0$ (i. e., $\sigma_p \to 0$ or $\sigma_c \to \infty$), then $E_N = 0$, which indicates that the initial two-mode quantum state of Eq. (2) becomes a separability quantum state; when $W$ increases, $E_N$ increases, which means the increasing of the entanglement of the initial two-mode quantum state. From Eqs. (16-19), we can also find that in the vacuum the evolution of the entanglement for the output two-mode quantum state becomes a constant, i.e., the logarithmic negativity $E_N = \frac{1}{2}\ln(1+4W^2)$ in the vacuum is independent of the propagating distances, similar to the property of the purity in the vacuum. But, due to the atmospheric turbulence, the value of $E_N$ may decrease and is strongly affected by the input parameters $\sigma_p$ ($\sigma_c$), $W$, and the atmospheric structure parameter $C_n^2$.

### III. Numerical analysis and discussion

Now we turn to analyze the propagation properties of the input two-mode spatially Gaussian-entangled quantum state in the atmosphere space in detail. In the following calculations, we take the wavelengths of two modes as $\lambda_1 = \lambda_2 = 632.8$ nm, and the transfer distances for two modes in two paths are the same with $z_1 = z_2 = z$, thus the maximal distribution distance $L$ between two modes is $L = 2z$.

Figure 2 shows the typical effect of the atmospheric turbulence on the evolutions of the fidelity, purity and logarithmic negativity (the entanglement). It is clear shown that the fidelity of the output two-mode quantum state decreases with the increasing of propagating distances; and as the atmospheric turbulence becomes stronger and stronger, the fidelity decreases faster and faster with the increasing distances [see the dot-dashed and solid lines for the cases when atmospheric turbulence is locally strong ($C_n^2 = 1\times 10^{-14}$ m$^{-2/3}$) and locally very weak ($C_n^2 = 1\times 10^{-17}$ m$^{-2/3}$)]. At the same time, from the inset of Fig. 2(a), one can find that even in the free space ($C_n^2 = 0$) the fidelity decreases quickly within the short-distance regions, mainly due to the light self-diffraction effect [i. e., the broadening of the probability density distribution, also see Eq. (8)]. For the property of the purity, denoting the degree of the mixedness of the output quantum state, from Fig. 2(b), it is clear seen that $\mu$ is always equal to one in the free space but decreases quicker and quicker in the stronger and stronger turbulent atmosphere. In the same way, for the evolution of the spatial entanglement, it is clear seen that the value of logarithmic negativity $E_N$ will decrease to be zero due to the effect of the atmospheric turbulence. In the very weak atmospheric turbulence ($C_n^2 = 1 \times 10^{-17} \, \text{m}^{-2/3}$), in this example, the entanglement could be preserved within the distances $z_1 = z_2 = z < 17.89$ km, i. e., the maximal distance of distributing quantum entanglement is about $L = 2z \cong 35.98$ km. In the vacuum ($C_n^2 = 0$), the logarithmic negativity is a constant [see the dot line in Fig. 2(c)]. With the increasing strength of the atmospheric turbulence, i. e., the increasing of the parameter $C_n^2$, the distance for preserving non-zero $E_N$ becomes shorter and shorter [see Fig. 2(c)]. For example, for the considerably strong atmospheric turbulence ($C_n^2 = 1 \times 10^{-14} \, \text{m}^{-2/3}$), the maximal distance is shortened to be about $L = 2z \approx 2 \times 0.86$ km. The result implies that the atmospheric turbulence strongly destroy the entanglement of the two-mode spatially entangled light fields and limit the transfer distance of the quantum entanglement, and the resulting output quantum state becomes a mixed state after passing through a distant atmospheric transfer.

Figure 3 shows the evolutions of the fidelity, purity and logarithmic negativity of the output two-mode quantum state inside the turbulent atmosphere in the cases of different values $W$ (i.e., with different input entanglement). It is clear seen that, see Figs. 3(a) and 3(b), with the fixed parameter $\sigma_p$, the fidelity decreases much faster with the increasing of the distances for the cases with larger values of $W$; and correspondingly the purity also decreases much faster for the cases with larger $W$. It indicates that both the purity and fidelity of the output two-mode quantum state are much fragile for the quantum state with much large $W$ when the parameter $\sigma_p$ is fixed. From Fig. 3(c), one can find that as the input entanglement increases, i. e., the increasing of $W$, the maximal transfer distance for keeping the output quantum state to be entangled gradually increases and is saturated to be an upper limit under the case with the fixed parameter $\sigma_p$. For example, in Fig. 3(c) the curves for $W = 10$ and $100$ are nearly overlapped except in the short-distance regions.

In Fig. 4, it shows the typical dependences of the fidelity, purity and logarithmic negativity for the cases that the parameter $\sigma_c$ (denoting the transverse spatial quantum correlation between two modes) of the input two-mode quantum state is fixed. From Fig. 4(a) and 4(b), one can find that when the parameter $W$ is optimally chosen, both the evolutions for the fidelity and purity of the output two-mode quantum state may have a slowest decreasing process with respect to the propagating distances for two modes. This indicates that with the fixed $\sigma_c$, in order to have the optimal transmissions for both the fidelity and purity of the output two-mode spatially entangled quantum state in the atmospheric channels, one has to choose a suitable input parameter $W$ (i.e., the input entanglement of the initial quantum state). However for the entanglement of the output quantum state, from Fig. 4(c), similar to Fig. 3(c), as the input entanglement increases (i. e., the increasing of $W$), the maximal transfer distance for preserving the entangled properties of the output two-mode quantum state also gradually increases and is saturated to be an upper limit. In this case with the fixed parameter $\sigma_c = 0.01\,\text{m}$, the upper limit of the maximal distributing distance of the quantum entanglement is about $L = 2z \approx 2 \times 37.9\,\text{km}$ for the sufficient large value of $W$. From Figs. 3(c) and 4(c), we have the following statement: *for the input two-mode spatially entangled quantum state with the fixed $\sigma_p$ (or $\sigma_c$), with the increasing of the input entanglement (i. e., the increasing $W$), the maximal transfer distance for keeping the entanglement of the output quantum state gradually increases and always has a maximal upper limit in a certain atmospheric condition.*

Figure 5 shows how the fidelity, purity and logarithmic negativity of the output two-mode quantum state depend on the changes of the input parameter $\sigma_p$ (or $\sigma_c$) and the propagating distances in the condition of the very weak atmospheric turbulence. Here we set the parameter $W$ unchanged, and then $\sigma_c$ varies as $\sigma_p$ changes (in order to keep $W$ as a constant). From Figs. 5(a) and 5(b), one can find that both the evolutions for the fidelity and purity of the output two-mode quantum state also decrease as the increasing of the propagating distances and have the optimal choices of $\sigma_p$ (or $\sigma_c$) to obtain the slowest decreasing process as functions of the propagating distances, which are similar to the effect in Fig. 4(a) and 4(b). In Fig. 5(c), the logarithmic negativity (entanglement) gradually decreases as the propagating distances increase, and may disappear after the sufficient long distances due to the atmospheric turbulence. From Fig. 5(c) and 5(d), one can find that for the initial two-mode spatially entangled quantum state with the fixed entanglement (i. e., $W$ is fixed), when $\sigma_p$ increases (i. e., $\sigma_c$ also increases), the maximal distributing distance for preserving the entanglement of the output quantum state in the atmospheric channel gradually increases, and with the suitable choice of the $\sigma_p$ there is a maximal upper limitation for the entanglement transfer; when $\sigma_p$ further increases, the maximal distributing distance of the entanglement decreases and becomes smaller and smaller. For example, in the case of $W=3$, the optimally maximal upper limitation $L \sim 2\times 36.6$ km occurs at $\sigma_p \approx 0.080$ m, while in the case of $W=1$, the optimally maximal upper limitation $L \sim 2\times 30.9$ km occurs at $\sigma_p \approx 0.045$ m. Therefore, we may have the conclusion that the propagation of the two-mode spatially entangled light fields is strongly affected by the atmospheric turbulence and also strongly depends on the input parameters of the initial two-mode quantum state. In the practical applications of the entanglement transmission, the input parameters of the initial two-mode quantum state have to be optimally chosen in order to have the high fidelity and the long-distance entanglement distribution.

## IV. Conclusion

In this work, we have investigated the propagation of two-mode spatially Gaussian-entangled quantum light fields passing through the turbulence atmosphere. The fidelity, purity and logarithmic negativity (entanglement) of the resulting quantum state via the long-distance atmospheric transportation are analytically derived. Using these derived formulae, we have analyzed in detail the effects of the atmospheric turbulences on the evolutions of quantum properties of the resulting two-mode quantum state under different input parameters of the initial two-mode quantum state. The main results show that the maximal distributing distance $L$ of quantum entanglement does strongly depend on the atmospheric turbulence and it becomes shorter and shorter for the stronger atmospheric turbulence, and both the fidelity and purity decrease much quicker as functions of propagating distances in the stronger atmospheric turbulence. Under a certain atmospheric condition, the input parameters of the initial two-mode spatially Gaussian-entangled quantum state also affect on the maximal distributing distance for preserving the entanglement and the evolutions of both the fidelity and purity of the output two-mode quantum state. The long-distance distribution of quantum entanglement becomes very important in the global quantum communication. Therefore we believe that our theoretical results are valuable for building the long-distance quantum information transfer and quantum imaging via free-space atmosphere link.

**ACKNOWLEDGEMENTS**

This work was supported by the National Nature Science Foundation of China (10604047) and by the financial support from RGC of HK Government (N_CUHK204/05, NSFC 05-06/01).

# FIGURE CAPTIONS

FIG. 1. Schematic of two-mode spatially Gaussian-entangled light fields through the paths 1 and 2 with the turbulent atmosphere. S denotes the source of the two-mode spatially entangled quantum state.

FIG. 2. Typical effects of the atmospheric turbulence on (a) fidelity $F$, (b) purity $\mu$ and (c) logarithmic negativity $E_N$, with different $C_n^2 = 1\times 10^{-14}\,\text{m}^{-2/3}$ (the dot-dashed line), $1\times 10^{-15}\,\text{m}^{-2/3}$ (the short-dashed line), $1\times 10^{-16}\,\text{m}^{-2/3}$ (the dashed line), $1\times 10^{-17}\,\text{m}^{-2/3}$ (the solid line), 0 (the dotted line). Note that in (a) both the solid and dotted lines are nearly overlapped.

FIG. 3. Changes of (a) the fidelity, (b) purity and (c) logarithmic negativity of the output two-mode quantum state through the turbulent atmosphere ($C_n^2 = 1\times 10^{-17}\,\text{m}^{-2/3}$) under different parameters $W$ and with the common parameter $\sigma_p = 0.01\,\text{m}$.

FIG. 4 Dependences of (a) the fidelity, (b) purity and (c) logarithmic negativity on the parameter $W$ and the propagating distance of the output two-mode quantum state through the turbulent atmosphere ($C_n^2 = 1\times 10^{-17}\,\text{m}^{-2/3}$), with the common parameter $\sigma_c = 0.01\,\text{m}$.

FIG. 5 Dependences of (a) the fidelity, (b) purity and (c-d) logarithmic negativity on the parameter $\sigma_p$ and the propagating distance for the initial two-mode quantum state, with the fixed parameter $W$ (i. e., the fixed entanglement), passing through the turbulent atmosphere ($C_n^2 = 1\times 10^{-17}\,\text{m}^{-2/3}$). In (a-c) we take $W = 3$ and in (d) we take $W = 1$.

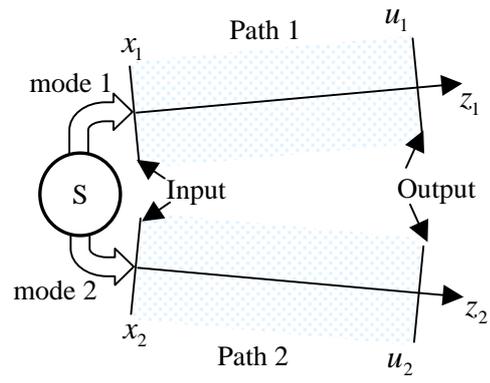

FIG. 1.

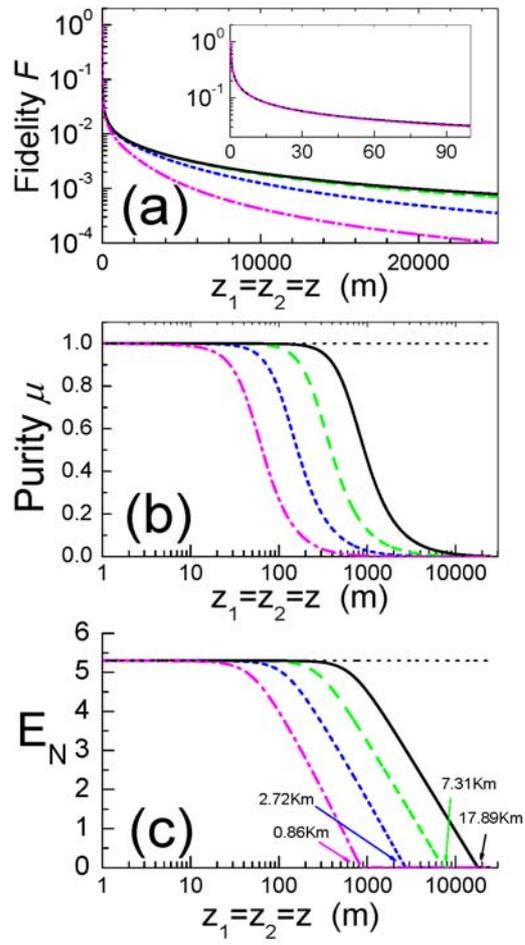

FIG. 2

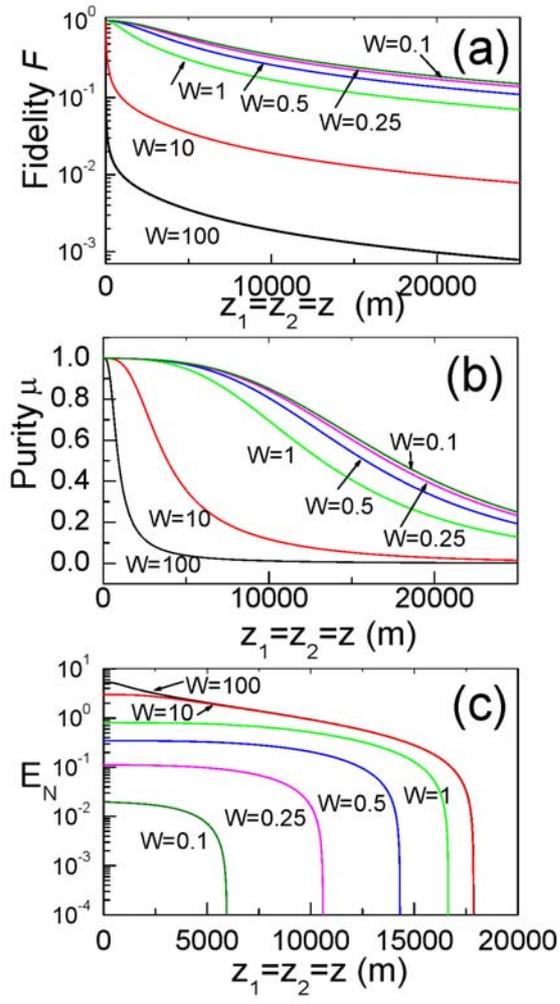

FIG. 3

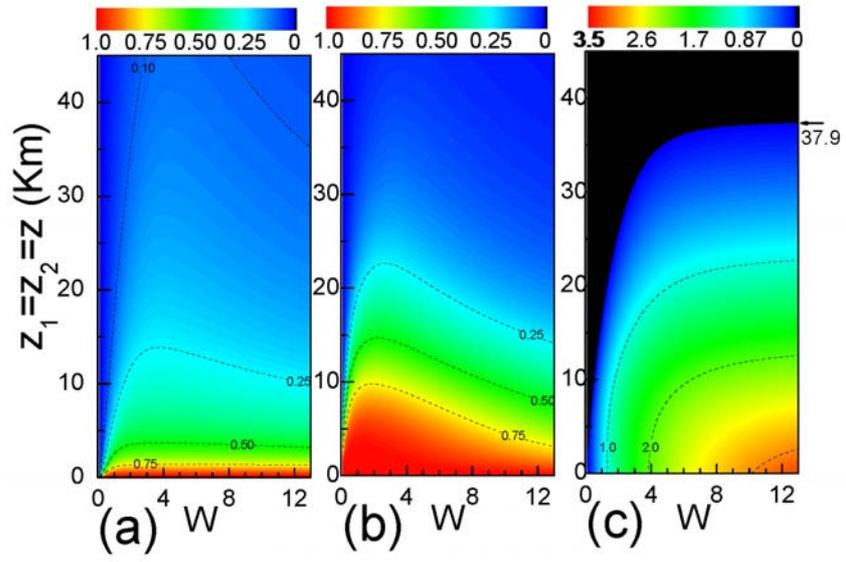

FIG. 4.

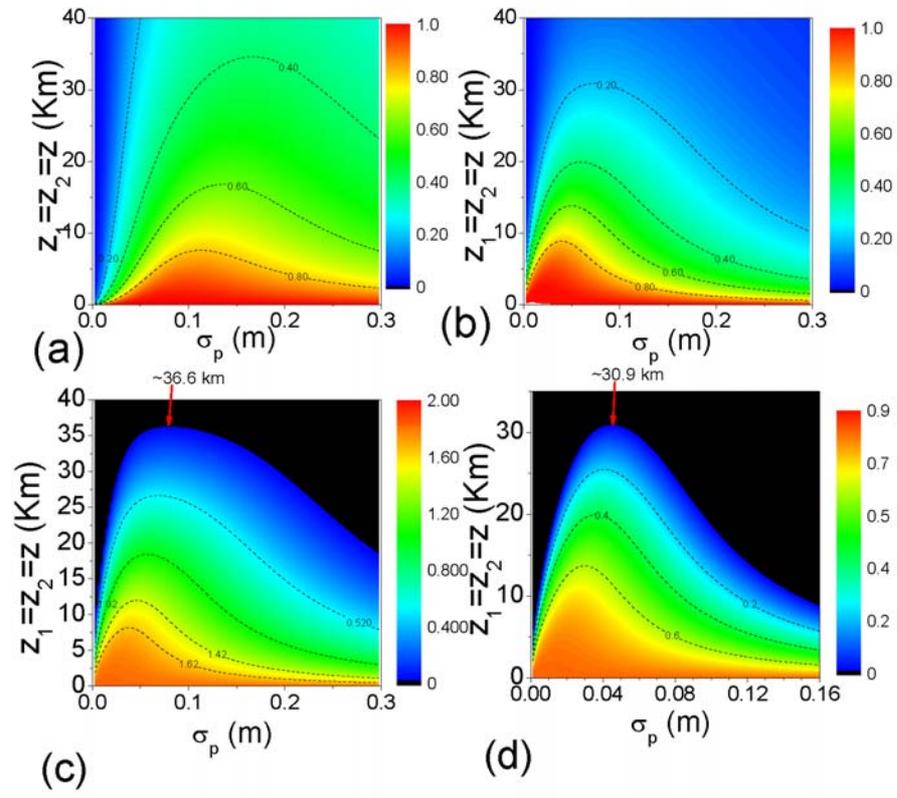

FIG. 5